\documentclass[pdflatex,sn-mathphys]{sn-jnl}

\jyear{2021}%

\usepackage{bussproofs}
\theoremstyle{thmstylethree}%
\newtheorem{theorem}{Theorem}
\numberwithin{theorem}{section}
\usepackage{xcolor}

\theoremstyle{thmstylethree}%
\newtheorem{corollary}[theorem]{Corollary}
\theoremstyle{thmstyletwo}%

\theoremstyle{thmstylethree}%
\newtheorem{definition}[theorem]{Definition}%

\theoremstyle{thmstylethree}%
\newtheorem{convention}[theorem]{Convention}%

\theoremstyle{thmstylethree}%
\newtheorem{lemma}[theorem]{Lemma}%

\newenvironment{bprooftree}
  {\leavevmode\hbox\bgroup}
  {\DisplayProof\egroup}

\begin{document}

\title[Effects of the Strict-Tolerant Approach on Constructive Logics]{Effects of the Strict-Tolerant Approach on Intuitionistic and Minimal Logic}





\abstract{This paper extends the literature on the strict-tolerant logical approach by applying its methods to intuitionistic and minimal logic. In short, the strict-tolerant approach modifies the usual notion of logical consequence by stipulating that, in order for an inference to be valid, from the truth of the premises must follow the non-falsity of the conclusion. This notion can also be generalized to define strict-tolerant metainferences, metametainferences and so on, which may or may not generate logics distinct from those obtained on the inferential level. It is already known that strict-tolerant definitions can make the notion of inference for non-classical logics collapse into the classical notion, but the strength of this effect is not yet fully known. This paper shows that intuitionistic strict-tolerant inferences also collapse into classical ones, but minimal ones do not. However, minimal strict-tolerant logic has the property that no inferences are valid (which is not carried over to the metainferential level). Additionally, it is shown that the logics obtained from intuitionistic, minimal and classical logic at at the metainferential level are distinct from each other.}

\author{Victor Barroso-Nascimento}
\author{German Mejia}




\maketitle







\section{Introduction}




The main goal of this paper is to further the study of non-classical logics by investigating some relations between criterias for inferential validity and constructivism. Specifically, we present technical results concerning a combination of the strict-tolerant approach \cite{StrictTolerant} with intuitionistic and minimal logic. It is claimed that, aside from making evident some perils of combining strict-tolerant notions of consequence with paraconsistent and constructive base logics, these results provides an answer to the questions raised by Fitting at \cite[pg. 393]{Fitting} of (i) whether there is such a thing as a strict-tolerant version of intuitionistic logic and (ii) what would happen if we were to define a strict-tolerant logic in which inferences are valid whenever from the intuitionistic truth of the premises follows the classical truth of one of the conclusions.

Our discussion is structured as follows. In the second section, we briefly comment on the relationship between the study of inferences and constructivism, providing some context for our technical results. In the third section, we offer the basic definitions required for our proofs. In the fourth section, we present this paper's main technical results. In the fifth and final section, we conclude by providing some brief comments on the results of section four, their perceived inner workings and some possible relations to future research.


\section{The inferential justification of constructivism}

Argumentative practices implicitly rely on the possibility of justifying claims by providing reasons for them. In the context of an argument, the act of drawing a conclusion from something purportedly providing it with a justification is called an \textit{inference}. Inferences are called \textit{valid} when they succeed in providing acceptable justifications and \textit{invalid} when they fail to do so. Since arguments themselves are structured as successive acts of inference, the acceptability of an argument depends on the acceptability of its inferences, so an argument is \textit{valid} whenever all its inferences are valid and \textit{invalid} when at least one is not.

Although the systematic study of argumentation has integrated logic since the works of Aristotle \cite{PrawitzValidityInference}, there is still no consensus on what makes inferences, and therefore arguments, valid. In fact, Prawitz singles this as the most fundamental question of General Proof Theory \cite{PrawitzFundamental}, the branch of logic dedicated to study of argumentative structures. It is also fundamental for philosophical doctrines in which inferences play a prominent role -- such as \textit{inferentialism}, according to which the contents of propositions are exhaustively determined by their possible uses as premises and conclusions of inferences \cite{BrandomInferentialism}.

The study of inferences and inferential validity is of particular interest in the context of constructive logics, for which inferentialism is contemporarily viewed as one of the most robust philosophical justifications. Michael Dummett and Dag Prawitz in particular have ostensively argued that inferences and the role they play in the determination of propositional meaning provides both a justification for the laws of intuitionistic logic and motives for rejecting classical canons of reasoning \cite[pgs. 245-279]{DummettMeta}\cite{DUMMETT19755} \cite{Placek1999} \cite{PrawitzLogicalConsequence}\cite{PrawitzMeaningProof}. Also worthy of note is the remarkable affinity between formal semantics defined using the concept of inference and intuitionistic logic \cite{sep-proof-theoretic-semantics}\cite{Sandqvist10.1093/jigpal/jzv021}\cite{Stafford10.1093/analys/anac100}\cite{GheorghiuLinear10.1007/978-3-031-43513-3_20}, although under suitable conditions the framework can also yield semantics for classical logic \cite{Sandqvist}\cite{gheorghiu2024prooftheoreticsemanticsfirstorderlogic}\cite{barrosonascimento2025prooftheoreticapproachsemanticsclassical}.

When it comes to logical consequence, traditional accounts establish truth-preservation, and sometimes relevance, as the main ingredient of validity \cite{StephenREad}. In such views, a valid argument must at least be capable of guaranteeing that from the truth of its premises follows the truth of its conclusion. Despite its simplicity, the feasibility of this requirement remains highly contentious \cite{MurziValidity}. It is also regarded as especially problematic by constructivists, as the very concept of truth -- unless given unorthodox epistemic readings instead of the traditional ontological one --  is considered inadequate for semantical analysis, either in general or specifically in the context of mathematics \cite{Prawitz2012}\cite{Martin-Lof1987-MARTOA}\cite{Dummett1998-DUMTFT}.


The use of different criteria for validity has been suggested in the literature, often as a direct answer to problems arising from the use of truth preservation. In particular, the literature on \textit{strict-tolerant logics} suggests that by adopting a \textit{weaker} criterion which submits premises to a ``strict” evaluation standard and conclusions to a ``tolerant” one it becomes possible to satisfactorily deal with problems pertaining to semantic paradoxes \cite{Cobrero}.  The approach stipulates that an inference is valid whenever from the truth of the premises follows at least the non-falsity of the conclusion, which is the same as truth-preservation in classical logic but yields a distinct entailment notion in other contexts \cite{StrictTolerant}\cite{Fitting}.

 As shown in \cite{BarrCarni} and \cite{bar}, this approach allows one to deal with paradoxes related to vagueness by semantically defining sequent-based logics in which the \textit{Cut} principle (transitivity of deduction) does not hold in general. Such systems easily allow one to provide interesting calculi for paraconsistent logics such as the Logic of Paradox \cite{Paradox}, whose affinity with the strict-tolerant approach had already been noted before \cite{StrictTolerant}. The literature also shows that, by applying the approach in a paraconsistent setting, it becomes possible to propose unorthodox solutions to traditional problems: in \cite{Hierarchy}, for instance, an hierarchy of paraconsistent logics formulated using the strict-tolerant approach is used to propose an entirely new criterion of identity between logics, although the philosophical significance of this hierarchy is challenged in \cite{Hlobil2022-HLOATS}.

 In light of the importance given to inferences in the philosophy of intuitionism, the investigation of how deviant concepts of inferential validity fare in constructive frameworks seems to be warranted. This paper will specifically investigate the inferential and metainferential behavior of strict-tolerant notions of validity in Kripke semantics for minimal and intuitionistic logic, the former being a paraconsistent version of the latter which arose from minor disagreements concerning the possibility of constructively justifying the principle of \textit{ex falso sequitur quodlibet} \cite[pg. 102]{Heyting}\cite{Johansson1937DerME}\cite{Molen}. The inclusion of minimal logic is justified not only by the proximity between both logics and their respective philosophical justifications, but also by the aforementioned well-known connection between paraconsistent logics and the strict-tolerant approach.

\section{Basic definitions}

 We start by defining semantics for intuitionistic and minimal logic, using the notation and definitions of \cite{Priest} and taking some elements from $\cite{MininimalSemantics}$.

\begin{definition}

The language $L$ of minimal and intuitionistic logic is comprised of atomic formulas, the unary logical connective $\bot$ and the binary logical connectives $\land$, $\lor$ and $\to$.

\end{definition}

\begin{convention}
$\neg A$ is used as an abbreviation for $A \to \bot$.
\end{convention}

\begin{definition}

A \textit{interpretation} is a triple (W, R, \textit{v}), in which W is a non-empty set of worlds $w$, $R$ is a reflexive and transitive relation on $W$ and $v$ is a function assigning one of two values $\langle 1,0 \rangle$ to pairs $\langle A, w \rangle$ comprised of formulas $A$ and worlds $w$.

\end{definition}

\begin{convention}
$v_w(A) = 1$ is used to express that the function $v$ assigns value $1$ to the pair $\langle A, w \rangle$, and $v_w(A) = 0$ to express the assignment of $0$ to the same pair.
\end{convention}

\begin{definition}

A \textit{minimal interpretation} is an interpretation in which the following constraint is satisfied:  for all $w$ and $w'$ in $W$, if $v_w(A) = 1$ and wRw', $v_{w'}(A) = 1$.

\end{definition}

\begin{definition}

A \textit{intuitionistic interpretation} is a minimal interpretation in which $v_w(\bot) = 0$ for all $w$.

\end{definition}

\begin{definition}

A \textit{classical interpretation} is a intuitionistic interpretation such that, for all $w$ and $w'$ in $W$ and all formulas $A$, $v_w(A) = v_{w'}(A)$. 

\end{definition}

\begin{definition}

The semantic clauses for molecular formulas in interpretations are as follows:

\begin{enumerate}
    \item  $v_w(A \land B) = 1$ if and only if $v_w(A) = 1$ and $v_w(B) = 1$.
    \item  $v_w(A \lor B) = 1$  if and only if $v_w(A) = 1$ or $v_w(B) = 1$.
    \item  $v_w(A \to B) = 1$ if for all w' such that wRw', $v_{w'}(A) = 0$ or $v_{w'}(B) = 1$
    
\end{enumerate}
\end{definition}

Notice that the accessibility relation $R$ becomes useless in classical interpretations, as the preservation of values across worlds boils the implicational clause down to making an implication true whenever the antecedent is false in all worlds or the consequent is true in all worlds (and thus, since the values of implications and conjunctions are also constant across worlds, the semantic collapses into usual two-valued Boolean truth functions).

\begin{definition}

Truth and falsity in an interpretation can be defined as followed:

\begin{enumerate}
    \item $A$ is \textit{true} in an interpretation $\langle W, R, v \rangle$ iff $v_w(A) = 1$ for all $w \in W$;
    
    \medskip
    
    \item $A$ is \textit{false} in an interpretation $\langle W, R, v \rangle$ iff $v_w(\neg A) = 1$ for all $w \in W$;
\end{enumerate}

\end{definition}

Three things should be noted about this definition:

\bigskip

\noindent (I) We are adopting a strong notion of falsity, according to which a formula is false if and only if its negation is true. This is intended to be a \textit{constructive} notion of falsity, so the falsity of a formula is only recognized when we have an actual refutation of it (and thus we are also able to assert its negation). Therefore, a non-true formula is not automatically false, which is very much in line with the spirit of intuitionistic and minimal logic;

\bigskip

\noindent (II) In accordance to what is usually expected of constructive theories, a formula can be neither true nor false in intuitionistic and minimal interpretations; 

\bigskip

\noindent (III) In intuitionistic logic, falsity of $A$ is equivalent to $v_w(A) = 0$ for all $w \in W$, which makes it so that a formula cannot be both true and false. This is not the case in minimal logic, which admits simultaneous truth and falsity (an unsurprising feature for a logic with paraconsistent characteristics). Another plausible definition for falsity would be to directly use the condition $v_w(A) = 0$ for all $w \in W$, but an unpleasant consequence of this would be that proofs of $\neg A$ would no longer establish the falsity of $A$ in minimal logic -- and thus this definition would change the meaning usually assigned to negation.

\begin{definition}
$\Gamma \vDash_m A$ holds if every minimal interpretation which makes all formulas in $\Gamma$ true also makes the formula $A$ true. 
\end{definition}

\begin{definition}
$\Gamma \vDash_i A$ holds if every intuitionistic interpretation makes all formulas in $\Gamma$ true also makes the formula $A$ true. 
\end{definition}

\begin{definition}
$\Gamma \vDash_c A$ holds if every classical interpretation  which makes all formulas in $\Gamma$ true also makes the formula $A$ true.
\end{definition}

We now proceed to the definition of strict-tolerant variants of those logics, drawing from the definitions in \cite{StrictTolerant} and \cite{Hierarchy}:

\begin{definition}
A strict-tolerant inference $\Gamma \Rightarrow \Delta$ holds if every interpretation which makes all formulas in $\Gamma$ true do not make all formulas in $\Delta$ false.
\end{definition}

\begin{corollary}
Let $\Delta = \{ A_1 , ... , A_n \}$ for non-empty $\Delta$. A strict-tolerant inference $\Gamma \Rightarrow \Delta$ holds under classical interpretations if and only if $\Gamma \vDash_c A$ holds, in which $A$ is the disjunction $A_1 \lor ... \lor A_n$.
\end{corollary}

Corollary 3.14 follows directly from classical bivalence, Definition 3.13 and the classical definition of disjunction. It shows that, unless we are interested in inferences with empty succedents, we need only consider classical inferences with at most a single formula on the succedent, as non-empty sets can be replaced by a single disjunctions without any loss.

The same can be shown for intuitionistic and minimal inferences:

\begin{theorem}
Let $\Delta = \{ A_1 , ... , A_n \}$ for non-empty $\Delta$. A strict-tolerant inference $\Gamma \Rightarrow \Delta$ holds under minimal (or intuitionistic) interpretations if and only if $\Gamma \Rightarrow A$ holds, in which $A$ is the disjunction $A_1 \lor ... \lor A_n$. 
\end{theorem}

\textbf{Proof}. For the left-to-right direction, let $\Gamma \Rightarrow \Delta$ hold. Then, by Definitions 3.13 and 3.9, for every interpretation which makes $\Gamma$ true we have $v_w(\neg A_k) = 0$ for some $A_k \in \Delta$ and some $w$ of the interpretation. For this $w$ it holds that there is a $w'$ with $w R w'$ such that $v_{w'}(A_k) = 1$ and $v_{w'}(\bot) = 0$. From $v_{w'}(A_k) = 1$ we can iterate the disjunction semantic clause to get $v_{w'}(A_1 \lor ... A_k ... \lor A_n) = 1$, and since $v_{w'}(\bot) = 0$ we have $v_{w'}( \neg (A_1 \lor ... A_k ... \lor A_n)) = 0$, which prevents $A_1 \lor .... \lor A_n$ from being false in the interpretation and thus establishes $\Gamma \vDash A$, as the result was proven for arbitrary interpretations which make $\Gamma$ true.

For the right-to-left direction, let $\Gamma \vDash A$. Then for every interpretation which makes $\Gamma$ true we have $v_{w}( \neg (A_1 \lor .... \lor A_n)) = 0$ for some $w$. Again, for this $w$ we have a $w'$ with $wRw'$ in which $v_{w'}( A_1 \lor ... \lor A_n) = 1$ and $v_{w'}(\bot) = 0$. By repeatedly decomposing the disjunction we have $v_{w'}(A_k) = 1$ for some $A_k \in \Delta$, and since $v_{w'}(\bot) = 0$ we have $v_{w'}(\neg A_k) = 0$, which prevents $A_k$ from being false. Since this can be done for any interpretation which makes $\Gamma$ true, we have $\Gamma \Rightarrow \Delta$. \qed

\bigskip

In order to simplify some of the proofs, from now on we implicitly use these results and consider only inferences with a single formula on the succedent.


\begin{definition}
The strict-tolerant intuitionistic consequence relation $\Gamma \vDash_{ST}^{i} A$ holds if $\Gamma \Rightarrow A$ holds for all intuitionistic interpretations.
\end{definition}

\begin{definition}
The strict-tolerant minimal consequence relation $\Gamma \vDash_{ST}^{m} A$ holds if $\Gamma \Rightarrow A$ holds for all minimal interpretations.
\end{definition}

We will now define the notion of \textit{strict-tolerant metainference}. A strict-tolerant metainference is a higher-order inference which has a (possibly empty) list of strict-tolerant inferences as its premises and a single strict-tolerant inference as its conclusions. We denote this relation by $\Rightarrow^{*}$, and define it as follows:

\begin{definition}
A strict-tolerant metainference $(\Gamma_1 \Rightarrow A_1), ..., (\Gamma_n \Rightarrow A_{n}) \Rightarrow^{*} (\Gamma_{n+1} \Rightarrow A_{n+1}) $ holds if every interpretation either (i) makes $\Gamma_k$ true and $A_k$ false for some k ($1 \leq k \leq n)$, (ii) does not make $\Gamma_{n+1}$ true or (iii) makes $A_{n+1}$ not false.
\end{definition}

In more intuitive but less practical terms, for a metainference to hold it must be the case that in every interpretation which makes $\Gamma_k$ true and $A_k$ not false (for all $\Gamma_k$ and $A_k$ on the antecedent, if any), if this interpretation makes $\Gamma_{n+ 1}$ true, it also makes $A_{n+1}$ not false. Notice also that condition (i) cannot be satisfied when the list of inferences on the antecedent is empty, which forces us to consider clauses (ii) and (iii) for all interpretations and thus makes validity of the metainference equivalent to validity of the inference on the succedent (compare clauses (ii) and (iii) with Definition 3.13).

\begin{definition}
Let $\Theta$ be a (possibly empty) sequence of strict-tolerant inferences and $S$ be a strict-tolerant inference. Then, the strict-tolerant minimal metaconsequence $\Theta \vDash^{m}_{STM} S$ holds if the metainference $\Theta \Rightarrow^* S$ holds in all minimal interpretations.
\end{definition}

\begin{definition}
Let $\Theta$ be a (possibly empty) sequence  of strict-tolerant inferences and $S$ be a strict-tolerant inference. Then, the strict-tolerant intuitionistic metaconsequence $\Theta \vDash^{i}_{STM} S$ holds if the metainference $\Theta \Rightarrow^* S$ holds in all intuitionistic interpretations.
\end{definition}

\begin{definition}
Let $\Theta$ be a (possibly empty) sequence  of strict-tolerant inferences and $S$ be a strict-tolerant inference. Then, the strict-tolerant classical metaconsequence $\Theta \vDash^{c}_{STM} S$ holds if the metainference $\Theta \Rightarrow^* S$ holds in all classical interpretations.
\end{definition}

Naturally, the notion of metainference can be generalized so as to create an hierarchy of logics, as done in \cite{Hierarchy}. However, as will be briefly shown, the inferential and meta-inferential levels are already sufficient to distinguish between classical, intuitionistic and minimal logics with regards to identity, so the meta-inferential level is sufficient for the purposes of this paper.

\section{Main results}

We now prove some preparatory lemmata and then proceed to prove this paper's main results.

\begin{lemma}[Glivenko's Theorem, Generalized]

If $\Gamma \vDash_c A$, then $\Gamma \vDash_i \neg \neg A$.
\end{lemma}

\textbf{Proof}. The standard proof can be seen in Proposition 2 of \cite{ONO2009246}, which shows that $\Gamma \vDash_{c} A$ implies  $\Gamma , \neg A \vDash_{i} \bot$ and allows us to conclude Lemma 4.1 through the deduction theorem for intuitionistic logic. \qed

\bigskip

This is a straightforward generalization of Glivenko's original theorem, which was proved only for empty $\Gamma$ \cite{Glivenko}. It essentially relies on the fact, when $A$ is a classical but not intuitionistic validity, every application of \textit{reductio ad absurdum} in the proof of $A$ can be transformed into a proof of negation with conclusion $\neg \neg A$. It may therefore also be viewed as a corollary of Seldin's normalization strategy, since it shows that if there is a classical proof of $A$ then there is a classical proof of $A$ containing only one application of \textit{reductio ad absurdum}, which always appears at the very last step \cite{Seldin}.

The result can be extended to the fragment of minimal logic without implication \cite{Ertola}. Surprisingly, if one does include implication in the language it no longer holds:

\begin{lemma}
There is a formula $A$ such that $\vDash_c A$ but $\nvDash_m \neg \neg A$.
\end{lemma}

\begin{corollary}
There is a set $\Gamma$ and a formula $A$ such that $\Gamma \vDash_c A$ but $\Gamma \nvDash_m \neg \neg A$.
\end{corollary}


\textbf{Proof.} We can obtain a quick semantic counterexample by considering an instance $\neg a \to (a \to b)$, with atomic $a$ and $b$, of the classical tautology $\neg A \to (A \to B)$. Let $I$ be minimal interpretation with only two worlds ($w$ and $w'$) such that: 

\bigskip

(I) $w R w'$ and the relations obtained through reflexivity and transitivity

of R hold, but not $w' R w$;

\medskip

(II) $v_w(a) = 1$ and $v_w(b) = v_w(\bot) = 0$;

\medskip

(III) $v_{w'}(a) = v_{w'}(\bot) = 1$ and $v_{w'}(b) = 0$. 

\bigskip

By Convention 3.2, clause 3 of Definition 3.8 and our choice of valuation function we have both $v_w(\neg a) = 0$ and $v_{w'}(\neg a) = 1$.

Since $v_{w'}(a) = 1$, $v_{w'}(b) = 0$ and $w'$ is related to itself via reflexitivity of R, the semantic clauses yield $v_{w'}(a \to b) = 0$, which together with $v_{w'}(\neg a) = 1$ yields $v_{w'}(\neg a \to (a \to b)) = 0$. However, since $v_{w'}(\bot)= 1$, we have both $v_{w'}(\neg (\neg a \to (a \to b))) = 1$ and $v_{w'}(\neg \neg (\neg a \to (a \to b))) = 1$.

Now, since wRw', we have $v_w(\neg a \to (a \to b)) = 0$ due to the valuation of $\neg a$ and $a \to b$ in $w'$. Since $v_w(\neg a \to (a \to b)) = 0$ and $v_{w'}(\bot) = 1$, as no other world is accessible from $w$, by clause 3 of definition 6 we also have  $v_w(\neg (\neg a \to (a \to b))) = 1$. However, since $v_w(\neg (\neg a \to (a \to b))) = 1$ and $v_w(\bot) = 0$, we have $v_w(\neg \neg (\neg a \to (a \to b))) = 0$, which makes it so that $\neg \neg (\neg a \to (a \to b))$ is not true in this interpretation (according to Definition 3.9), and so $\nvDash_m \neg \neg (\neg a \to (a \to b))$ by Definition 3.10. \qed

\bigskip

We will now proceed to prove the first main result of his paper. The proof is quite general and relies heavily on Glivenko's Theorem, and the distinction between its status in minimal and intuitionistic logic may be taken as an explanation for the discrepant results that will be presented for both.

\begin{theorem}
$\Gamma \vDash_{STM}^{i} A$ if and only if $\Gamma \vDash_c A$. 
\end{theorem}

\textbf{Proof.} The left-to-right direction follows immediately from the fact that the set of all classical interpretations is a subset of the set of all intuitionistic interpretations. $\Gamma \vDash_{STM}^{i} A$ implies that no intuitionistic interpretation makes all the formulas of $\Gamma$ true and $A$ false, which also implies that no classical interpretation makes all the formulas in $\Gamma$ true and $A$ false. Since classical interpretations collapse into usual truth-functional semantic, we immediatelly have $\Gamma \vDash_c A$.

Now for the right-to-left direction. Assume, for contradiction, that for some $\Gamma$ and $A$ we have $\Gamma \vDash_c A$ and $\Gamma \nvDash_{ST}^{i} A$. By Lemma 4.1 we have $\Gamma \vDash_i \neg \neg A$, and so every intuitionistic interpretation which makes the formulas in $\Gamma$ true also make $\neg \neg A$ true. Since $\Gamma \nvDash_{ST}^{i} A$, by Definitions 3.13 and 3.16, there must be an intuitionistic interpretation $I$ which makes all formulas in $\Gamma$ true but makes $A$ false. By Definition 3.9, $A$ is false in a intuitionistic interpretation if and only if $\neg A$ is true. Then, both $\neg A$ and $\neg \neg A$ are true in $I$ and thus receive value $1$ in all its worlds, and so $\bot$ must also receive value $1$ in all $w$ due to the semantic clause for implication. But $\bot$ cannot receive the value $1$ in intuitionistic interpretations. Contradiction. Thus, if $\Gamma \vDash_c A$ then $\Gamma \vDash_{ST}^{i} A$. \qed

\bigskip

The structure of this proof also makes it evident that it is not restricted to the characterization of intuitionistic logic via Kripke models, as no feature of those models is used in an essential fashion. Since it relies only on Glivenko's Theorem and the definition of falsity of a formula in an interpretation, every semantic for intuitionistic logic in which falsity of $A$ is defined as equivalent to truth of $\neg A$ will experience this collapse. The particular relevance of this is made evident when one points out that  some logicians denounce the inadequacy of Kripke models for formalizing intuitionistic semantics \cite{Wagner}, which renders a proof which does not depend on the inner workings of such models especially interesting. It is completely inessential that our result uses a intuitionistic definition of the notion of ``truth" formulated in Kripke models, as any notion of ``construction" in which falsity/refutability of $A$ were to be equated with a ``construction" of $\neg A$ and truth/provability of ``A" were to be equated with a ``construction" of $A$ would face the same issue.  

We claim this result provides a strong, general, negative answer to the question raised by Fitting in \cite[pg. 393]{Fitting} of whether there is such a thing as an intuitionistic strict-tolerant logic, as any combination of the strict-tolerant definitions (if tolerant truth is equated with non-falsity) with semantics for intuitionistic logic which equate truth of $\neg A$ with falsity of $A$ will collapse into classical logic at the inferential level. The negative results also apply to Fitting's proposal \cite[pg. 393]{Fitting} of a strict-tolerant logic in which the conclusion of an inference is required to be classically true whenever the premises are intuitionistically true: classical truth of $A$ is equivalent to classical truth of $\neg \neg A$, which in turn is equivalent to the non-falsity of $A$, leading us back to the same strict-tolerant definition of inference and thus to the same issues.


To conclude commentaries on our mains results regarding intuitionistic logic, we point out that Glivenko's theorem can also be combined with intuitionistic proofs of equivalence between $\neg A$ and $\neg \neg \neg A$ to obtain his second theorem, $\vDash_c \neg A \Leftrightarrow \ \vDash_i \neg A$,  which we may also generalize using the procedure of Lemma 4.1 to obtain $\Gamma \vDash_c \neg A \Leftrightarrow \Gamma \vDash_i \neg A$. These results bind both negations very closely together and present many challenges to their technical (and perhaps even conceptual) separation. In fact, since in many cases they are prone to collapse \cite{Binder}, many mixed systems -- such as those used in ecumenical approaches to logic \cite{ClassicInt} \cite{Marin} -- do not distinguish between classical and intuitionistic negation at all. It is evident, then, that strict-tolerant definitions can only be used together with intuitionistic logic without producing collapses if one either does not use non-falsity when evaluating the consequent of strict-tolerant inferences or promote a significant overhaul of the intuitionistic concepts of falsity or negation.

We now proceed to prove results concerning minimal logic.

\begin{theorem}
$\Gamma \vDash_{ST}^{m} A$ implies $\Gamma \vDash_c A$.
\end{theorem}

\textbf{Proof}. Trivial. We use the same reasoning as in the left-to-right direction of Theorem 4.4, noting that both the set of all intuitionistic interpretations and the set of all classical interpretations are subsets of the set of all minimal interpretations. \qed

\begin{theorem}
There is a $\Gamma$ and an $A$ such that $\Gamma \vDash_c A$ but $\Gamma \nvDash_{STM}^{m} A$. 
\end{theorem}

\textbf{Proof.} Let $\Gamma$ be the empty set and let formula $A$ be the classical tautology $\neg A \to (A \to B)$. Consider now the minimal interpretation used in the proof of Lemma 4.2 and Corollary 4.3.. According to Definition 3.9, $\neg ( \neg a \to (a \to b))$ is true in this interpretation, as we have both $v_w(\neg ( \neg a \to (a \to b))) = 1$ and $v_{w'}(\neg ( \neg a \to (a \to b))) = 1$. Likewise, Definition 3.9 makes $\neg a \to (a \to b)$ false in this interpretation, and so $\nvDash_{ST}^{m} \neg a \to (a \to b)$. \qed

\bigskip

It is particularly interesting to notice that, in many traditional sources (such as Heyting's original monograph \cite{Heyting}), $\neg A \to (A \to B)$) is precisely the axiom taken from intuitionistic logic in order to obtain minimal logic.

In light of those basic results, one would naturally be led to ask what kind of logic is characterized by the combination of strict-tolerant inferences with minimal logic. Is it a paraconsistent logic? Does it have partial classical behavior? How much intuitionistic behavior it retains?

The answer, it turns out, is very surprising:

\begin{theorem}
 For any $\Gamma$ and $A$, $\Gamma \nvDash_{ST}^{m} A$.
\end{theorem}

\textbf{Proof}. Consider the interpretation with a single world $w$ in which every atom is true, including $\bot$. Definition 3.8 can then be used to prove inductively that every molecular formula is also true. But, according to Definition 3.9, this means that all formulas in $\Gamma$ are true and $\neg A$ is true, and so $A$ is false. Thus, $\Gamma \nvDash_{ST}^{m} A$. \qed

Even though Theorem 4.6 was proved by considering the structure of a concrete minimal interpretation, it is actually a vacuous result. As such, the logic obtained by applying the strict-tolerant method to minimal logic has no valid inferences, much like the logic TS analyzed in \cite{BarrCarni}.

Fortunately, these results do not carry over to the metainferential level, in which we indeed have valid consequences. Consider, for example, the following metainferential rule, in which the inference above the line is taken to be the antecedent of Definition 3.18 (that is, the sequence of inferences to the left of $\Rightarrow^*$) and the inference below the line is taken to be its succedent (that is, the inference to the right of $\Rightarrow^*$):

\bigskip

\begin{center}
\begin{bprooftree}
\AxiomC{$\Rightarrow A$}
\UnaryInfC{$\Rightarrow A$}
\end{bprooftree}
\end{center}

\bigskip

 The interpretation considered in Theorem 4.7, Lemma 4.2 and Corollary 4.3 cannot be used to invalidate metainferences containing inferences with a non-empty set of premises: since it makes every formula both true and false, it invalidates any such inferences on the antecedent of the metainference, thus making the metainference itself hold vacuously in this interpretation by satisfying clause (i) of Definition 3.18. For all other interpretations it is evident that, conditional on $A$ not being false, $A$ is not false, and the results follows from the mere existence of such an interpretation (consider, for example, the interpretation containing a single world $w$ with $v_w(A) = 1$ but $v_w(\bot) = 0$)).


In the context of the identity criteria proposed in \cite{Hierarchy}, even though the logics obtained from classical  and intuitionistic logic by the strict-tolerant method cannot be distinguished at the inferential level, this level is already sufficient to make a distinction between them and miminal logic. From the distinction at the inferential level we also immediately obtain distinction at the metainferential level, as validity of metainferences having an empty set of premises (that is, containing an empty sequence of inferences on its antecedent) is reducible to validity of the inference on the succedent, as briefly noted in the commentary presented immediatelly after Definition 3.18. However, we still need to investigate what happens in the meta-inferential level of intuitionistic logic, since it could be the case that it cannot be differentiated from classical logic.

\begin{theorem}
For some $\Theta$ and some $S$ we have $\Theta \vDash_{STM}^{c} S$ but $\Theta \nvDash_{STM}^{i} S$ and $\Theta\nvDash_{STM}^{m} S$.
\end{theorem}

\textbf{Proof.} Consider the following metainference, in which the inferences of $\Theta$ stand above the line and $S$ stands below:

\bigskip

\begin{center}
 
\begin{bprooftree}
\AxiomC{$\Rightarrow A$}
\AxiomC{$\Rightarrow B$}
\BinaryInfC{$\Rightarrow A \land B$}
\end{bprooftree}

\end{center}

\bigskip

It is easy to verify that this metainference is valid in the case of classical logic, as any classical interpretation making the conclusion of both inferences occurring above the line non-false makes both $A$ and $B$ true and thus also $A \land B$ true (hence also non-false). In intuitionistic and minimal logic, however, it is not valid -- which, as will be shown, essentially follows from the partial failure of DeMorgan's Laws, as in both logics we have $\neg (A \land B) \nvDash \neg A \lor \neg B$.

Let $I$ be a minimal interpretation with three worlds $(w, w'$ and $w''$). For this interpretation, consider the following specifications, with $a$ and $b$ atomic:

\bigskip

(I) Only $wRw'$, $wRw''$ and the relations obtained through reflexivity and

transitivity of R hold;

\medskip

(II) $v_w(a) = v_w(b) = v_w(\bot) = 0$;

\medskip

(III) $v_{w'}(a)  = 1$ and $v_{w'}(b)  = v_{w'}(\bot) = 0$; 

\medskip

(IV) $v_{w''}(a) = v_{w''}(\bot) = 0$ and $v_{w''}(b)  = 1$; 

\bigskip

We can use the semantic clauses to show that neither $a$ nor $b$ are false in this interpretation, as $v_{w'}(\neg a) = 0$, $v_{w''}(\neg b) = 0$ and falsity requires assignment of $1$ to the negation in all worlds. As such, this interpretation validates the semantic inferences $\Rightarrow a$ and $\Rightarrow b$. However, in no world of this interpretation we have both $v(a) = 1$ and $v(b) = 1$, which makes $a \land b$ false in the interpretation. Thus, since this atomic instance of the metainference is invalid, this metainference in general is invalid in minimal logic. Furthermore, since this interpretation is both a minimal and a intuitionistic interpretation, it is also invalid in intuitionistic logic. \qed

To conclude this section, it is worth noticing that the logics defined at the metainferential level by intuitionistic and minimal logic have a particular paraconsistent flavour, allowing us to draw comparisons  similar to those in \cite{BarrCarni} concerning validation and invalidation of distinct formulations of the principle of explosion at the inferential and metainferential level. 

From Theorems 4.4 and 4.7 it follows that $A, \neg A \Rightarrow B$ is valid on the system obtained from intuitionistic logic but not on the one obtained from minimal logic, even though they seem to have similar metainferential behaviour.  Moreover, consider the following two metainferences:

\bigskip

\begin{center}
    \begin{bprooftree}
    \AxiomC{$\Rightarrow \bot$}
    \UnaryInfC{$\Rightarrow B$}
    \end{bprooftree}
    \qquad \qquad
        \begin{bprooftree}
    \AxiomC{$\Rightarrow A$}
    \AxiomC{$\Rightarrow \neg A$}
    \BinaryInfC{$\Rightarrow B$}
    \end{bprooftree}
\end{center}

\bigskip

Both logics validate the left one, but not the right one.

Since $\neg \bot$ is defined as $\bot \to \bot$, it will receive value $1$ at all worlds $w$ regardless of the value of $\bot$, and so $\bot$ is false in all minimal and intuistionitic interpretations -- which makes the inference on the left hold vacuously. However, for a counterexample to the second metainference, consider an interpretation with three worlds $w$, $w'$ and $w''$, with $wRw'$ and $wRw''$ but neither $w'Rw''$ nor $w''Rw'$ and in which $\bot$ and $b$ receives value $0$ at all worlds and $a$ receives value $1$ only at $w'$. Since $v_{w'}(\neg a) = 0$, $a$ is not false, and since $v_{w''}(\neg a) = 1$ and $v_{w''}(\bot) = 0$ we have $v_{w''}(\neg \neg a) = 0$, and thus $\neg a$ is also not false. However, since $b$ receives value $0$ at all worlds, $\neg b$ also receives value $1$, and so $b$ is false.

Since Theorem 4.8 also establishes a meaningful difference between those logics and the ones defined by the metainferences of classical logic, it follows that those are indeed some brand new logics with paraconsistent features. The proof of Theorem 4.8 also hints that the constructive nature of the underlying definitions has concrete effects on the notion of metainferential validity, as the particular metainference we considered seems to fail essentially due to the intuitionistic/mininal failure of one of the directions of the DeMorgan equivalences.

\section{Conclusion}

The results presented in this paper show that some particular features of intuitionistic and minimal systems interact with strict-tolerant definitions in unexpected ways. Glivenko's theorems and the proximity between classical and intuitionistic negation makes it so that intuitionistic strict-tolerant inferences collapse into classical inferences, and minimal logic's feature of allowing trivial models able to validate any premise and invalidate any conclusion makes it so that every minimal strict-tolerant inference is invalid. However, those features are partially eliminated when we go from the inferential to the metainferential level. Even though some behaviours observed at the inferential level are carried over to the metainferential one, metainferences provide enough wiggle room for those systems to escape their respective pathologies. As such, we are still able to obtain interesting logics from those combinations.

Our use of Glivenko's theorems in the results for intuitionistic logic makes them highly general, as every semantic definition in which falsity of $A$ is equated with truth of $\neg A$ will face similar issues. Since this is the expected reading of intuitionistic negation, this might be taken as a proof that the task of providing an intuitionistically acceptable strict-tolerant logic is effectively untenable.  On the other hand, the failure of DeMorgan's laws in intuitionistic and minimal logic impacts what those logics validate at the metainferential level, preventing the total collapse into either classical or a trivial logic.

Intuitively, the results for intuitionistic logic show that, in light of the equivalence between truth of classical negation and truth of intuitionistic negation, the further equivalence between falsity and truth of negation makes logics defined through recourse to intuitionistic or classical falsity prone to a specific kind of collapse. Since classical truth of $A$ is also equivalent to classical falsity of $\neg A$, the definition does not even have to reference falsity explicitly, which explains why the collapse happens both in strict-tolerant intuitionistic logic and in logics applying intuitionistic truth standards to the premises of an inference and classical truth standard to its conclusion.  Moreover, the results for minimal logic show that logics in which truth and falsity may simultaneously obtain are prone to a different kind of collapse when we use them to define logics that essentially rely on distinctions between the two truth values.



In short, our work provides both a positive and a negative result to the literature, showing that the combination of strict-tolerant definitions with semantics for constructive logics are problematic but not entirely so. It also shows that strict-tolerant criterias of inferential validity cannot be taken to justify intuitionistic or minimal logic. Furthermore, the general nature of the proofs we present provide a cautionary tale: one should be very careful of mixing the strict-tolerant approach with logics that allow trivial models or which have a negation similar to that of classical logic, as those combinations are particularly prone to collapses.





\bigskip


\bigskip






\bibliography{sample}


\end{document}